\documentclass[twocolumn,epjc3]{svjour3}

\usepackage{amsmath}
\usepackage{graphicx}
\usepackage{color}

\newcommand{\zpred}{$z\mathrm{^{max}_{pred}}$}
\newcommand{\ztrue}{$z\mathrm{^{max}_{true}}$}

\journalname{Eur. Phys. J. C}

\begin{document}

\title{A Data-Directed Paradigm for BSM searches:\newline
the bump-hunting example}

\author{Sergey Volkovich\thanksref{addr1}
        \and
        Federico De Vito Halevy\thanksref{addr1}
        \and
        Shikma Bressler\thanksref{addr1,e1}
}

\thankstext{e1}{Corresponding author: shikma.bressler@cern.ch}

\institute{Department of Particle Physics \& Astrophysics, Weizmann Institute of Science, Rehovot, Israel \label{addr1}}

\date{Received: date / Accepted: date}

\maketitle


\begin{abstract}
We propose a data-directed paradigm (DDP) to search for new physics. Focusing on the data without using simulations, exclusive selections which exhibit significant deviations from known properties of the standard model can be identified efficiently and marked for further study. Different properties can be exploited with the DDP. Here, the paradigm is demonstrated by combining the promising potential of neural networks (NN) with the common bump-hunting approach. Using the NN, the resource-consuming tasks of background and systematic uncertainty estimation are avoided, allowing rapid testing of many final states with only a minor degradation in the sensitivity to bumps relative to standard analysis methods.
\end{abstract}

\textbf{Introduction.} Despite its success in describing the elementary particles and their interactions, the Standard Model (SM) is still incomplete \cite{Wein18}. Many models extending beyond the Standard Model (BSM) have been developed over the years predicting the existence of new resonances. Thus, the search for new resonances, either theoretically-predicted or model-agnostic, is a core strategy for discovery in experimental high energy physics (e.g. recently \cite{ATLA20a,ATLA19a,ATLA20b}). 

With almost no exception\footnote{In \cite{ATLA19b,CMS2021nyc,CDF:2008voc}, discrepancies between the data and Monte Carlo prediction were searched for in a large variety of final states.}, all BSM searches have been conducted following the blind analysis paradigm, in which an enormous amount of time and effort is invested before looking at the data, i.e. on background modeling and systematic uncertainty estimation. These resource-intensive tasks have allowed only a limited region in the space spanned by all observables (``observable-space") to be explored to-date. Indeed, searches typically focus on inclusive final states --- di-lepton, di-photon, di-jet, etc. --- ignoring all other observables and avoiding exclusive selections such as di-lepton + jets, di-jets + missing transverse momentum, di-photon within a $t\Bar{t}$ topology, etc. Moreover, within the studied final states, event selection is usually optimized relative to predefined signal models. So far, no significant indication of BSM physics has been found.

Complimentary to the blind analysis paradigm, we propose a data-directed paradigm (DDP) which begins by efficiently identifying regions of interest in the data. Similarly to \cite{ATLA19b,CMS2021nyc,CDF:2008voc}, albeit without using MC simulation, the strategy consists of quickly searching the observable-space for exclusive regions exhibiting a significant deviation from some fundamental SM property. Such regions should be considered for data-directed signal hypotheses and further examined using traditional analysis techniques. Like in \cite{ATLA20b}, no Monte-Carlo (MC) simulation is used. Thus, the search is not sensitive to MC mismodelling or limited MC statistics. Given the large number of plausible signals which could manifest in an infinite number of exclusive regions, and moreover, the limited time, manpower and resources at hand, searches like the proposed DDP might provide our best chance at discovering BSM physics.

\textbf{A Data-Directed Paradigm.} A DDP search can be realized with two key ingredients:
\begin{enumerate}
    \item A theoretically well-established property of the SM with respect to which deviations can be searched for — here we exploit the fact that within the SM, in absence of resonances, almost any invariant mass distribution is smoothly falling. Other properties of the SM, such as flavour symmetry \cite{Bres14} or forward-backward symmetry could also be exploited once detector effects are taken into account (as implemented for instance in \cite{ATLA17}).
    \item An efficient algorithm to scan the observable-space in search for deviations — here we train a deep neural network (NN) to map any invariant mass distribution into a distribution of statistical significance for excesses of events (“bumps”). The latter is known as a “$z$” distribution and is based on the profile likelihood ratio test for positive signals \cite{Cowa11}. Different algorithms should be developed when searching for deviations from other SM properties.
\end{enumerate}
The challenge of bump-hunting is an excellent showcase for a search in the DDP\footnote{In \cite{CDF:2008voc}, bump-hunting was exploited by comparing data to MC prediction in many different final states.}; even a simple implementation achieves good accuracy. As long as the underlying background distribution is smoothly falling, a single trained NN, as described in this letter, can quickly perform statistical inference from many selections of observed data. For example, when adapting it to a narrower mass range, it predicted a maximum significance in agreement with the di-muon results presented in \cite{ATLA19a} in seconds. Crucially, it avoids the time- and effort-consuming tasks of full background and systematic uncertainty estimation currently carried out for every invariant mass distribution under consideration. This way, a potentially unlimited quantity of exclusive distributions can be scanned and large regions in the observable-space can be covered. Nevertheless, event-by-event optimization for bump enhancement, as studied for instance in \cite{Fari20,Coll19,Coll18}, is left to future work.

Bumps identified using the DDP are likely to be caused by statistical fluctuations. These will disappear when tested with more data. Bumps originating from systematic uncertainties due to detector effects (trigger thresholds, kinematic edges, etc.) should appear in Monte Carlo simulations as well and can be ruled out. Among the bumps found which do not disappear with added data and do not appear in simulation, the most significant ones should be considered as BSM signal hypotheses and devoted a dedicated analysis. Inevitably, some may be due to mismodelled systematic effects.

\textbf{A neural network implementation.}
The NN we employ is trained in a supervised manner, for which we generate a set of artificial training and testing data. These contain inputs which simulate realistic distributions in observed data (in contrast to individual events, as in \cite{Fari20,Coll19,Coll18}), and can be further tailored for any given search. Inputs are matched to analytically calculated $z$ distributions as targets. When given an invariant mass distribution, the NN predicts a $z$ distribution which shows where and how likely it is that the data contains a bump. Once the NN has been trained, we validate that its predictions are consistent and that its loss value converges. Finally, its predictions are evaluated on the test set and we discuss its performance.

We generate the inputs of the NN as 100 bin histograms of observed events, $N=B+S$. These are representative of data with high statistics and dynamic range (the bin width reflecting a given detector resolution). The generation process is illustrated in Fig. 1. Each input is composed of a smoothly decaying background curve, $B$, to which Poisson fluctuations are added, and a localized Gaussian signal, $S$, whose significance is defined relative to the fluctuated background. We calculate bin-by-bin the corresponding NN target, $z$, which we use to approximate the significance distribution given the unfluctuated background and assuming a Gaussian signal shape \cite{Cowa11}. Each input and target pair is collectively referred to as a ``sample". All samples are globally scaled to the interval $[0,1]$ under a linear transformation before being utilized by the NN. 
%
\begin{figure}
\centering
\includegraphics[width=0.7\linewidth]{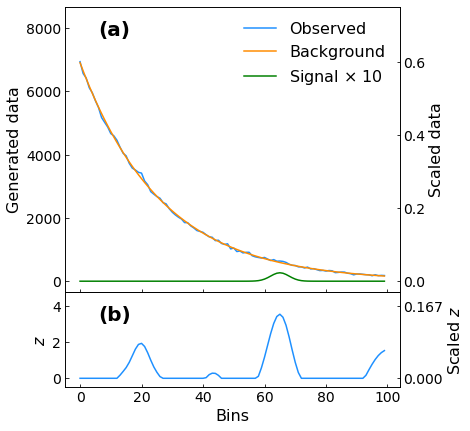}
\caption{Illustration of the sample generation procedure. (a) A smoothly decaying background curve (\textit{orange}) is generated over 100 bins. Each bin is assigned a Poisson fluctuation. A signal with a significance relative to the fluctuated background (\textit{green}) is added to it, producing the observed data (\textit{blue}). (b) The corresponding significance distribution, $z$, is calculated analytically. The left and right axes in both panels show the non-scaled and scaled distributions, respectively.}
\label{fig:1}
\end{figure}

A variety of smoothly falling backgrounds is modelled by randomly selecting one of the following ten functional forms for each sample:
%
\begin{equation*} \label{eq1}
\begin{split}
&be^{-ax},\ \ ax+b,\ \ \frac{1}{ax}+b,\ \ \frac{1}{ax^2}+b,\ \ \frac{1}{ax^3}+b,\\
&\frac{1}{ax^4}+b,\ \ a\left(x-x_2\right)^2+y_2,\ \ -a\cdot\ln\left(x\right)+b,\\
&\left(y_1-y_2\right) \cos \left(a\left(x-b\right)\right)+y_2,\ \ \cosh\left(a\left(x-x_2\right)\right)+b.
\end{split}
\end{equation*}
\noindent
The parameters $a$ and $b$ are defined such that each curve decays between two points, $\left(x_1,y_1\right)$ and $\left(x_2,y_2\right)$, where $x_1 < x_2$ are the centers of the extreme bins and $y_1 > y_2$ are randomized from the interval [100,10000].

Gaussian shaped signals are generated with mean values distributed randomly between bin 25 and bin 76. The width (standard deviation) of the signals is fixed at 3 bins. To improve the desired feature detection, the NN is trained with a data set containing signals with significance in the range [1,20]$\sigma$. The performance of the NN is determined on a testing data set by evaluating its ability to identify bumps with a significance of 3$\sigma$ --- the common definition of a ``hint'' for BSM physics.

Various NN architectures can be used. Here, we choose an architecture based on a dense layer followed by six 1-dimensional convolutional layers. The latter are intended for feature-detection, while the former is useful in suppressing position-dependent biases. A ``rectified linear unit" activation function is used. The ``Adam" optimizer is used to minimize the ``mean squared error” loss function over 200 epochs at a learning rate of 0.0003 with a batch size of 100. We generate a total of 600,000 training samples, 20\% of which are used for validation, and 150,000 testing samples.

\begin{figure}
\centering
\includegraphics[width=0.65\linewidth]{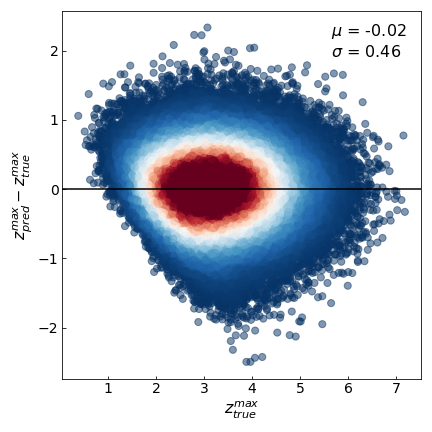}
\caption{The difference between \zpred{} and \ztrue{} as a function of \ztrue{}. Dense regions are shown in red (roughly corresponding to the $1\sigma$ region), while sparse regions are shown in blue.}
\label{fig:2}
\end{figure}

\textbf{Results.}
The accuracy of the NN prediction is illustrated in Fig.~\ref{fig:2} in terms of the difference between the maximal predicted significance, \zpred{}, and the one calculated via the profile likelihood ratio test, \ztrue{}. All generated test samples are included in the figure; in over 87\% of these the predicted peak was found within 1 bin of \ztrue{}. A mean ($\mu$) of $-0.02$ indicates a negligible bias in the prediction and a $0.46$ standard deviation ($\sigma$) measures its precision. The asymmetry seen as a sharp edge in the third quadrant originates from the small number of maximal $z$ predictions below one.
%

\begin{figure}
\centering
\includegraphics[width=0.7\linewidth]{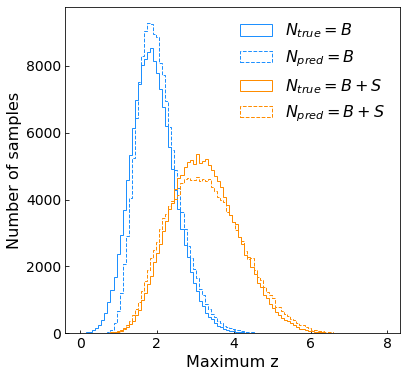}
\caption{The distribution of \ztrue{} (solid line) and \zpred{} (dashed line) for samples with no signal added (blue) and for samples with a 3$\sigma$ significance signal added (orange).}
\label{fig:3}
\end{figure}

We are interested in finding samples with bumps of 3$\sigma$ significance while rejecting samples without bumps. Fig.~\ref{fig:3} shows \ztrue{} in a solid line and \zpred{} in a dashed line for samples with no signal added (blue) and for samples with a 3$\sigma$ significance signal added (orange). In a traditional bump-hunting search, the signal significance is evaluated relative to an estimated background. Thus, the measured significance of a 3$\sigma$ signal could fluctuate around this value. This is the origin of the width of the \ztrue{} distributions: the signal is generated with a significance relative to the fluctuated background and its \ztrue{} is evaluated relative to the smooth background. 
%

\begin{figure}
\centering
\includegraphics[width=0.7\linewidth]{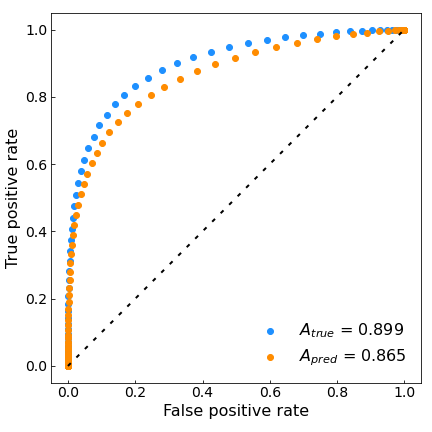}
\caption{True (blue) and predicted (orange) ROC curves and their associated areas under curve, $A_{\mathrm{true}}$ and $A_{\mathrm{pred}}$.}
\label{fig:4}
\end{figure}

According to the Neyman-Pearson lemma (see e.g.~\cite{Lehm05}), \ztrue{} provides the most powerful signal to background separation. It relies on exact knowledge of both the background and signal shapes. Yet, despite using no a priori knowledge of the two, the signal to background separation in \zpred{} is only slightly degraded relative to \ztrue{}. This is quantified in terms of receiver operating characteristic (ROC) curves in Fig.~\ref{fig:4}, obtained from the distributions of Fig.~\ref{fig:3}. The true (blue) and predicted (orange) ROC curves show the efficiency to correctly identify a 3$\sigma$ bump versus the false positive rate of selecting samples with no injected bump. The area under the true curve, $A_{\mathrm{true}}$, is 0.899 while the area under the predicted curve, $A_{\mathrm{pred}}$, is 0.865, which implies a degradation in performance of less than $4$\%. In other words, the probability that based on the NN output a selection will be marked as potentially interesting approaches the probability that a traditional method would do the same. 
%

We also confirmed that the NN is able to generalize in identifying with comparable accuracy bumps over linear combinations of the background forms (Eq.~\ref{eq1}), and over an unseen 10\textsuperscript{th} shape when trained on 9 background shapes\footnote{This test was carried out for the background shapes $be^{-ax}$ and $\frac{1}{ax^4}+b$.}. Thus, it is unrestricted by specific background forms in its capacity to detect bumps, which goes beyond the potential of traditional techniques. Similar performance was obtained in additional scenarios: when testing on distributions with lower and higher statistics (in the ranges between 100-500 and 5000-10000, respectively), when extending the bump region from the bin range 25-76 to 5-96, and when training and testing on samples with wider bumps, of either 4 or 5 bins.

\textbf{Validation.} We validate the convergence of the loss value achieved by increasing either the number of epochs or the size of the training data set. In terms of $A_{\mathrm{pred}}$ from Fig.~\ref{fig:4}, the NN performance varies insignificantly, by less than 1\% when moving past 200 epochs (100,000 input samples) or 500,000 input samples (100 epochs).

Consistency was ensured by comparing the NN predictions in two scenarios (with 100,000 training samples and 100 epochs). First, we trained four different NNs using an independent training data set for each and compared their predictions on a common testing data set (with 25,000 samples). Second, the performance of each of the NNs was separately compared on four different testing data sets. In all cases, the accuracy when separating signal from background was unaffected.

\textbf{Discussion.} We have presented a data-directed paradigm, complementary to the blind analysis paradigm, and demonstrated one of its possible implementations using the concept of bump-hunting. We have shown that a NN can be trained to efficiently identify bumps over smoothly falling backgrounds without being given any a priori information about the background or the bump's position. Relative to the most powerful test statistic (profile likelihood ratio), which relies on exact knowledge of both the background and signal shapes, the performance of the NN was only inferior by less than 4\% when considering the area under the ROC curve. Since for each data distribution the NN prediction is obtained within a couple of seconds (compared to a year or more when following the blind analysis paradigm), these results pave the way towards scanning the overwhelming observable-space that is being measured in experiments searching for bumps. Examples could be searches for di-lepton, di-jet, di-photon, jet-lepton-missing $E_T$ resonances in events containing, in addition, any other set of objects.

In the search for BSM physics we must leave no stone unturned. Complementary to traditional theory-directed blind analysis searches, the DDP should be pursued as well. With the expected ramp up of the Large Hadron Collider, existing data should be thoroughly explored. A first milestone could be demonstrating sensitivity to bumps in regions already investigated. If needed, dedicated NNs could be trained to account for scenarios not covered by the current implementation (e.g. different dynamic ranges, bins or widths) and other architectures could be explored. The search for bumps over smoothly falling backgrounds is just one example of a property of the SM that could be considered. Others such as flavour symmetry \cite{Bres14} or forward-backward symmetry could be exploited as well. Given the challenge ahead, searches like the proposed DDP might provide our best chance at discovering BSM physics.


\paragraph{Declarations}
\textbf{Funding} This work was supported by Grant No. 2871/19 from the Israeli Science Foundation (ISF), Grant No. I-1506-303.7/2019 from the German Israeli Foundation (GIF) and by the Sir Charles Clore Prize.
\newline
\textbf{Conflicts of interest / Competing interests} The authors have no conflicts of interest to declare that are relevant to the content of this article.
\newline
\textbf{Availability of data and material} The data sets generated and analysed during this work are available from the corresponding author on reasonable request.
\newline
\textbf{Code availability} The code of this work is available from the corresponding author on reasonable request.



\begin{thebibliography}{4}

\bibitem{Wein18}
S.~Weinberg, Phys. Rev. Lett. \textbf{121}, 220001 (2018)

\bibitem{ATLA20a}
ATLAS Collaboration, JHEP \textbf{03}, 145 (2020)

\bibitem{ATLA19a}
ATLAS Collaboration, Phys. Lett. B \textbf{796}, 68-87 (2019)

\bibitem{ATLA20b}
ATLAS Collaboration, Phys. Rev. Lett. \textbf{125}, 051801 (2020)

\bibitem{ATLA19b}
ATLAS Collaboration, Eur. Phys. J. C \textbf{79}, 120 (2019)

\bibitem{CMS2021nyc}
CMS Collaboration, Eur. Phys. J. C \textbf{81}, 629 (2021) 

\bibitem{CDF:2008voc}
CDF Collaboration, Phys. Rev. D \textbf{79}, 011101 (2009)

\bibitem{Bres14}
S.~Bressler, A.~Dery, A.~Efrati, Phys. Rev. D \textbf{90}, 015025 (2014)

\bibitem{ATLA17}
ATLAS Collaboration, Eur. Phys. J. C \textbf{77}, 70 (2017)

\bibitem{Cowa11}
G.~Cowan, K.~Cranmer, E.~Gross, O.~Vitells, Eur. Phys. J. C \textbf{71}, 1554 (2011)

\bibitem{Fari20}
M.~Farina, Y.~Nakai, D.~Shih, Phys. Rev. D \textbf{101}, 075021 (2020)

\bibitem{Coll19}
J.~Collins, K.~Howe, B.~Nachman, Phys. Rev. D \textbf{99}, 014038 (2019)

\bibitem{Coll18}
J.~Collins, K.~Howe, B.~Nachman, Phys. Rev. Lett. \textbf{121}, 241803 (2018)

\bibitem{Lehm05}
E.L.~Lehmann, J.P.~Romano, \textit{Testing Statistical Hypotheses}, 3rd edn. (Springer, New York, 2005)


\end{thebibliography}
\end{document}